\documentclass[11pt,a4paper]{article}
\usepackage[left=20mm, top=20mm, right=20mm, bottom=20mm, nohead=5mm, nofoot=5mm]{geometry}
\usepackage[utf8]{inputenc}
\usepackage{authblk}
\usepackage{indentfirst}
\usepackage{misccorr}
\usepackage{graphicx}
\usepackage{amsmath}
\usepackage{amssymb}
\usepackage{multicol}
\usepackage{bm}
\usepackage{longtable}
\usepackage{pdflscape}
\usepackage{epstopdf}
\usepackage{multirow}
\usepackage{adjustbox}
\usepackage{amsfonts}
\usepackage{fancyhdr}
\usepackage{ifthen}
\usepackage{fancyheadings}
\usepackage{setspace}
\usepackage{xcolor}
\usepackage[round,authoryear,comma,sort&compress,numbers]{natbib}
\usepackage[toc,title,page]{appendix}
\usepackage[hidelinks]{hyperref}
\usepackage{url}

\hypersetup{
	colorlinks=true,
	linkcolor=green,
	citecolor=blue,
	filecolor=magenta,
	urlcolor=cyan,
	pdftitle={Sharelatex Example},
	bookmarks=true,
	pdfpagemode=FullScreen,
}
\fancypagestyle{plain}{\chead{\texttt{\textcolor{brown}{Communications of BAO, Vol. 67, Issue 2, 2020, pp. 232-235}}}}
\title{\textbf{INT monitoring survey of Local Group dwarf galaxies: star formation history and chemical enrichment}}
\author[1,2]{T. Parto\thanks{tahereparto@gmail.com, Corresponding author}}
\author[3]{Sh. Dehghani}
\author[1,4]{A. Javadi}
\author[1]{E. Saremi}
\author[5]{J. Th. van Loon}
\author[1]{H. Khosroshahi}
\author[2]{M. T. Mirtorabi}
\author[2]{H. Abdollahi}
\author[2]{M. Gholami}
\author[4]{S.A. Hashemi}
\author[1]{M. Navabi}
\author[6]{M. Noori}
\author[1]{S. T. Aghdam}
\author[6]{M. Torki}
\author[2]{M. Vafaeizade}

\affil[1]{\scriptsize School of Astronomy, Institute for Research in Fundamental Sciences (IPM), PO Box 19395-5531, Tehran, Iran}
\affil[2]{\scriptsize Physics Department, Alzahra University, Vanak, 1993891176, Tehran, Iran}
\affil[3]{\scriptsize  Physik. Institut, University of Cologne, Zulpicher Str. 77, D-50937 Cologne, Germany}
\affil[4]{\scriptsize Department of Physics, Sharif University of Technology, Tehran, 11155-9161, Iran}
\affil[5]{\scriptsize  Astrophysics Group, Lennard-Jones Laboratories, Keele University, Staffordshire ST5 5BG}
\affil[6]{\scriptsize Physics Department, Faculty of Science, University of Zanjan, Zanjan 45371-38791, Iran}

\begin{document}
	\pagestyle{empty}
	\newpage
	\pagestyle{fancy}
	\label{firstpage}
	\date{}
	\maketitle

\begin{abstract}
The Local Group (LG) hosts many dwarf galaxies with diverse physical characteristics in terms of morphology, mass, star formation, and metallicity. To this end, LG can offer a unique site to tackle questions about the formation and evolution of galaxies by providing detailed information. While large telescopes are often the first choices for such studies, small telescope surveys that perform dedicated observations are still important, particularly in studying bright objects in the nearby universe. In this regard, we conducted a nine epoch survey of 55 dwarf galaxies called the “Local Group dwarf galaxies survey” using the 2.5m Isaac Newton Telescope (INT) in the La Palma to identify Long-Period Variable (LPV) stars, namely Asymptotic Giant Branch (AGB) and Red Super Giant (RSG) stars. AGB stars formed at different times and studying their radial distribution and mass-loss rate can shed light on the structure formation in galaxies. To further investigate the evolutionary path of these galaxies, we construct their star formation history (SFH) using the LPV stars, which are at the final stages of their evolution and therefore experience brightness fluctuations on the timescales between hundred to thousand days. In this paper, we present some of the results of the Local Group dwarf galaxies survey.

\end{abstract}
	
	
\section{Introduction}
Dwarf galaxies are the most common type of galaxies in the universe. The importance of the internal and external processes (e.g., supernova explosion; interaction with the massive halos) on the evolution of these small galaxies are well known, though with many unanswered aspects (\citealt{2014ApJ...789..147W}; \citealt{2020ApJ...894..135S}). The star formation history (SFH) is a robust tracer of how different internal or external mechanisms affect the evolution of a galaxy (\citealt{2019IAUS..344..125S}).

The cool asymptotic giant branch (AGB) stars with luminosities of $\sim$ $10^4 L_{\odot}$ and wide age ranges, from $100$ Myr to older than $10$ Gry, are well-known probes of the stellar population of galaxies in the near-infrared (\citealt{AGB}; \citealt{2011MNRAS.411..263J}, \citeyear{2017MNRAS.464.2103J})

especially in the nearby universe. Many evolved AGBs, to be specific, thermally pulsing AGBs (TP-AGB), are long-period variables (LPVs; \citealt{2017ApJ...835...77M}), therefore they experience brightness fluctuations on the timescales between 100 to 1300 days due to the low surface gravity (\citealt{2019IAUS..343..283J}). TP-AGBs are responsible for a significant fraction of the integrated light of a galaxy, they also contribute meaningfully to the chemical enrichment of the interstellar medium (ISM) through different mechanisms that drive the stellar wind. Red supergiant (RSG) stars with the look-back time $10^7$ years are another major dust producers and are examples of LPVs and their brightness variates between 600 to 900 days (\citealt{2019IAUS..343..283J}).

This survey aims to uniformly determine the SFH of 55 dwarf galaxies of the Local Group (LG) by adopting a novel method. This method was first proposed and applied to the M33 by (\citealt{8214023}) and relies on the identification of LPVs. There are some other successful applications of this method on dwarf galaxies in LG (e.g., IC 1613 \citealt{2019MNRAS.483.4751H}; LMC and SMC \citealt{2014MNRAS.445.2214R}; NGC 147 and NGC 185 \citealt{2017MNRAS.466.1764H}).
Using this method, We can also monitor the amount of produced dust by LPVs, and investigate their role in the star formation and evolution of a galaxy. Surveying this large sample of LG dwarfs, enables us to determine the evolutionary dependense of dwarfs galaxies on the enviroment, such as proximity to the host galaxies and compare it with internal effects like the stellar mass or gas content (\citealt{2019IAUS..339..336S}).

%The first paper of this survey (Saremi 2020) explain the survey completely and we refer you to read it.
%We determined the amplitude and mean brightness of LPVs 

\section{Observations and Data}
	
The sample of dwarf galaxies studied in this survey consists of all observable Andromeda system of satellites in the northern hemisphere, along with 20 satellites of Milky Way and some isolated and transitional dwarfs (\citealt{Saremi_2017}). We exclude galaxies that have been studied with Javadi's method before. 
	
The Observations were made in 9 epochs between 2015-2018 using the Wide Field Camera (WFC), an optical mosaic camera on the INT telescope. WFC consists of four $2048 \times 4096$ CCDs, with a pixel scale of 0.33 arcsec/pixel.
	
We used the i-band filter that is suited for identifying dusty AGBs. We also performed observations in the V-band to obtain the changes in the color (temperature) of stars. For each galaxy, the observation nights are separated by a month or more to identify LPV stars.
	
We reduce the WFC images using the THELI image processing pipeline (\citealt{2005AN....326..432E}). After the reduction process, We perform photometry for the crowded-field using DAOPHOT II (\citealt{1987PASP...99..191S}) to construct a catalog for each galaxy. Furthermore, a completeness test carried out using the ADDSTAR task in the DAOPHOT II package to show the depth of our photometry.
To remove the foreground contamination, we impose selection criteria on the proper motion and parallax of stars estimated in the $GAIA$ Data Release 2 (\citealt{2018A&A...616A...1G}). The complete description of observations and details of the photometry procedure is available at (\citealt{2020ApJ...894..135S}).

	%0% completenes
\section{Method}
	
\subsection{LPV candidates detection}
	
\label{sec:Method}
To identify LPV candidates, we employed a method described in (\citealt{1996PASP..108..851S}) to determine the variability index L for each star. Then we estimated a threshold for variability, using the variability distribution for the stars in different magnitude bins (\citealt{2011MNRAS.411..263J}; \citealt{2020ApJ...894..135S})

With insufficient observation night, we can not obtain a meaningful period for LPVs. However, there is a correlation between the amplitude and period of a variable star (\citealt{2019ApJ...877...49G}), and large amplitude variables (LAV) are usually evolved AGBs at the late stage of their life.
	
\begin{figure}[ht]
		\centering
		\includegraphics[width=0.7\textwidth]{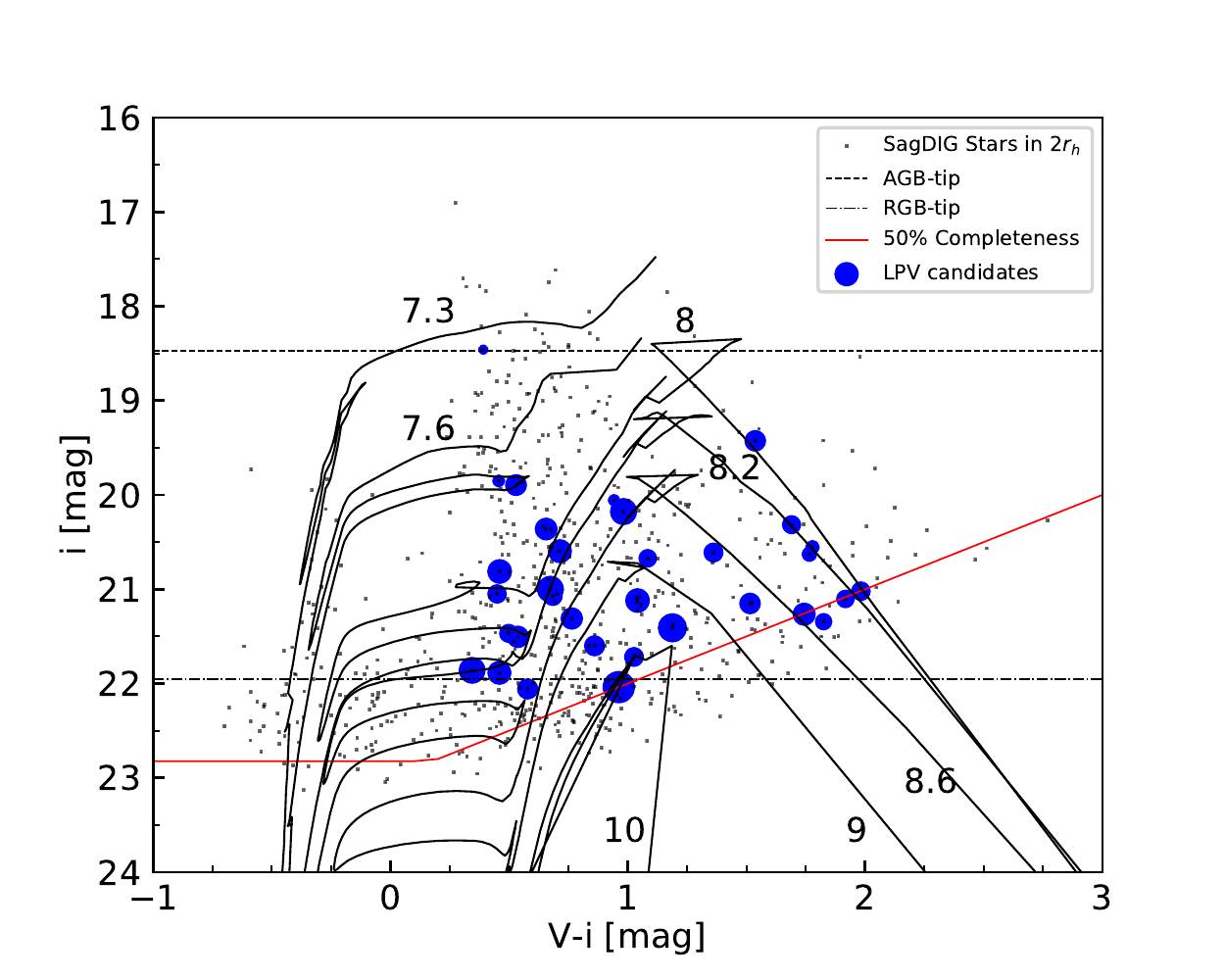}
		\caption{[i] vs. [V-i] CMD for SagDIG with the overplotted isochrones. Black dots denote the stars within the two half-light radius of this galaxy. Blue circles are the LPV candidates with a size scaled to their amplitude. The dotted line marks RGB-tip and AGB-tip. The $50\%$ completeness is represented by the solid red line.
			\label{fig:SagDIG}
		}
	\end{figure}

Fig. \ref{fig:SagDIG} shows the color-magnitude diagram (CMD) of SagDIG, one of the dwarf irregular galaxies in our sample with the overplotted PARSEC–COLIBRI isochrones (\citealt{2017ApJ...835...77M}). LPV candidates are shown as blue circles, with a size scaled to their amplitude in the i-band with values between $0.1-1.87$ $mag$. 

%The amplitude of LPVs increases towards the redder with decreasing of brightness.
	
%LPV's amplitude increase as their magnitude increase.(elham)
%Amplitudes increases towards the...
%de reddening

\subsection{SFH from LPV pulsation}
The luminosity of AGB stars reaches a maximum at the final stage of life, hence can be used to estimate the birth mass of the star. For determining the mass, we construct the mass function of LPVs (for the suitable metalicity) by interpolating mass-luminosity relation using the PARSEC–COLIBRI isochrones. After estimating age of LPVs using the mass-age relation, we determine pulsation duration by fitting multiple Gaussian functions to the mass-pulsation values in the isochrones that show strong pulsations.

The star formation rate (SFR) function $\xi(t)$ introduced in (\citealt{8214023}) takes mass, age and pulsation duration of LPVs and estimates the stellar mass formed per year ($M_{\odot}y^{-1}$):

\begin{equation}
	\xi(t) = 
	\frac{\mathrm{d  n(t) } }{\delta t}  
	\frac{ \intop\nolimits_{min}^{max} f_{IMF}(m)m\:dm }{ \intop\nolimits_{m(t)}^{m(t+dt)} f_{IMF}(m)m\:dm}.
	\label{eqn:2}
\end{equation}
	
where $f_{IMF}$ is the initial mass function (\citealt{2001MNRAS.322..231K}), $dt$ represents different age bins, and $dn$ is the number of observed LPVs in each age bins.

	% It is possible to reduce the size of a figure among other changes (see the instructions).  Here is an example:
	% \articlefigure[width=.5\textwidth]{example.jpg}

\section{Results and On–going works}
%structure infall time reinization , ...
Here we present results of published papers for three galaxies: And I and And VII, two dwarf spheroidal and satellite of Andromeda, and IC 10, an irregular and isolated galaxy.
%(consclusion + discussion + future owrk)

(\citealt{2020ApJ...894..135S}) detected 59 LPV candidates within the two half-light radii of And I, including five extreme AGBs (X-AGB).
They also modeled the spectral energy distribution (SED) of these stars, using $DUSTY$ code (\citealt{1997MNRAS.287..799I}) and mid-IR bands measurement from Spitzer (3.6 and 4.5 $\mu m$), WISE (W1=3.4, W2=4.6, W3=11.6, and W4=22 $\mu m$) (\citealt{2014yCat.2328....0C}; \citealt{2010AJ....140.1868W}), INT near-infrared i-band catalog and SDSS (u, g, r ,z) filters. They evaluated the total mass-loss rate of $3.5 \times 10^{-5} M_{\odot} yr^{-1}$ from five X-AGBs and thirteen dusty AGBs, which suggest low growth of stellar mass ($\sim$ 10 $\%$) in AND I in the next 10 Gyr.

(Navabi et al. Submitted, 2020) detected 43 LPV candidates within the half-light radii of And VII, and estimated the SFR peak of this area about 0.002 $M_{\odot}yr^{-1}$ at Z=0.0007. Probably, And VII was quenched by environmental impacts after infall into Andromeda’s virial radius (\citealt{2020svos.conf..383N}, \citeyear{mahdie-ipm}).

(\citealt{2019IAUS..344...70G}) detected around 10000 AGB stars in IC 10 in the area of CCD4 ($\sim$ 0.07 square-degree). They found that the AGB population concentrated more in the central region of IC 10, while the red giant branch (RGB) and RSG stars are more spread (\citealt{mahtab2020}).

In the following, we complete this study for other dwarfs and construct the dust map of the galaxy. We later discuss our results in light of different structure formation scenarios and the importance of internal feedbacks in LG dwarf galaxies.

\clearpage % To force this stuff to happen by this point in the text, otherwise these will probably end up after the references.

\section*{\small Acknowledgements}
\scriptsize{The observing time for this survey is provided by the Iranian National Observatory and the UK-PATT allocation of time to programs I/2016B/09 and I/2017B/04 (PI: J. van Loon). The authors thank the Iranian National Observatory and the School of Astronomy (IPM) for the financial support of this project.}

\bibliographystyle{ComBAO}
\nocite{*}
\bibliography{references}

\newpage
\appendix
\renewcommand{\thesection}{\Alph{section}.\arabic{section}}
\setcounter{section}{0}

\end{document}